\documentclass[RNAAS, modern]{aastex62}
\begin{document}

\shorttitle{\textit{Kepler K2} Analysis of EU Cnc}
\shortauthors{Hill et al.}

\title{The Intriguing Polar EU Cancri in the Eyes of \textit{Kepler K2}}

\author{Katherine Hill}
\affiliation{Bishop McGuinness Catholic High School, Oklahoma City, OK 73118 USA}

\author{Colin Littlefield}
\affiliation{Department of Physics, University of Notre Dame, Notre Dame, IN 46556 USA}

\author{Peter Garnavich}
\affiliation{Department of Physics, University of Notre Dame, Notre Dame, IN 46556 USA}

\author{Paula Szkody}
\affiliation{Department of Astronomy, University of Washington, Box 351580, Seattle, WA 98195 USA}

\correspondingauthor{Colin Littlefield}
\email{clittlef@alumni.nd.edu}

\keywords{stars:individual (EU Cnc); novae, cataclysmic variables; white dwarfs; accretion, accretion disks}

\section{Introduction}

EU Cnc is a polar, a binary star system in which mass is transferred from a companion star to a strongly magnetized white dwarf, generally emitting X-rays and cyclotron radiation due to the release of gravitational potential energy. The system is a proper-motion member of M67 \citep{williams13}, an open cluster 4.0 $\pm$ 0.5 Gy old \citep{perci03}. The Gaia distance to EU Cnc has a large error bar \citep[$770^{+740}_{-410}$~pc;][]{bailer18} but is consistent with the 883 $\pm$ 1~pc mean distance to the stars in M67 reported by \citet{gao18}.

Campaign 16 (C16) of the \textit{Kepler K2} mission continuously observed the system for 80 days at the one-minute short cadence and Campaign 18 (C18) did so again for 51 days at the thirty-minute long cadence. These observations are significant because EU Cnc is one of just four polars observed by the \textit{Kepler} spacecraft.

\begin{figure}
    \centering
    \includegraphics{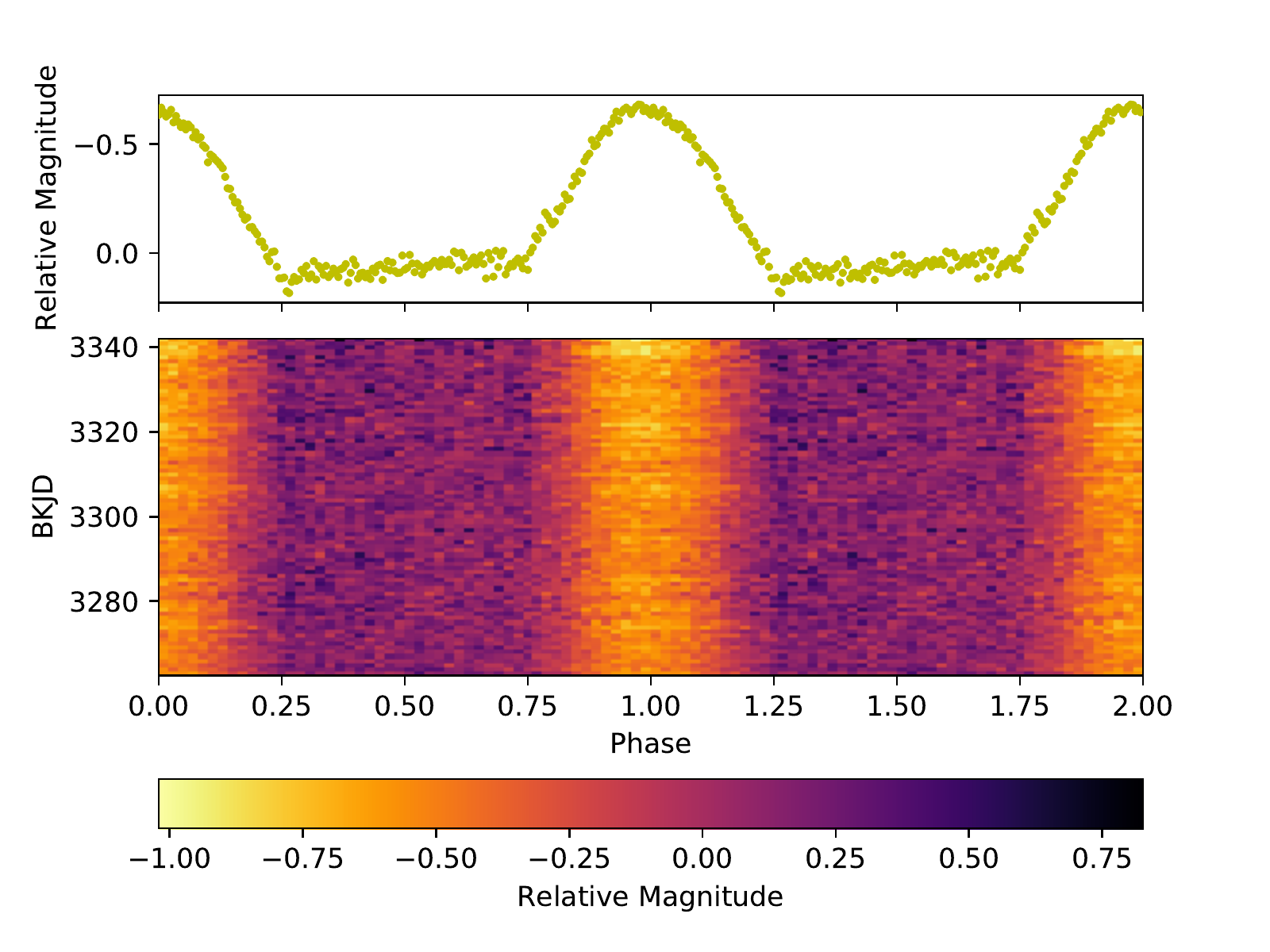} 
    \caption{{\bf Top:} Phase-averaged orbital light curve of EU Cnc. {\bf Bottom:} 2-D light curve. The faint horizontal bands are due to imperfect background subtraction due to spacecraft pointing jitter and the presence of a nearby bright star. There is no evidence of significant flickering, indicating a remarkably stable mass transfer rate. There are weak dips at phases 0.25 and 0.75 of unknown origin.}
    \label{fig:lightcurve}
\end{figure}

\section{Observations}

We extracted the C16 and C18 light curves using {\tt lightkurve} \citep{lightkurve}. We selected large extraction apertures that minimized the effect of spacecraft pointing jitter by collecting as much of the light from EU~Cnc as possible while also minimizing contamination from a nearby bright star. We sigma clipped the light curve to eliminate extreme outliers and applied a background subtraction due to the high background in both target-pixel files. Due to the significantly lower time resolution of the C18 light curve, we focus the bulk of our analysis on the C16 light curve.


\section{analysis}

The light curves in C16 and C18 are extremely faint (an average flux of $\sim$30 e$^{-}$ s$^{-1}$) but show noisy, period humps every 2.1 hours. We created a Lomb-Scargle power spectrum to determine the system's period, which we measured to be $0.087065\pm0.000002$~d. We interpret this to be the orbital period of the system. In addition to showing three additional orbital harmonics, the C16 power spectrum contains several sampling-related aliases, the strongest of which occurs at 440.4~c~d$^{-1}$, equivalent to exactly three-tenths of the short-cadence sampling rate. This alias is surrounded by a series of upper and lower sidebands, the spacing of which (48.9~c~d$^{-1}$) is equivalent to the long-cadence sampling frequency. These high-frequency signals are therefore instrumental in origin.

In the upper panel of Fig.~\ref{fig:lightcurve}, we phase the C16 light curve to the orbital period and average it into non-overlapping bins, each with a phase width of 0.005 to extract a smooth, almost noiseless orbital waveform. This waveform displays a sharp, hump-like pulse and a faint region, each lasting for half of the orbit. We attribute this behavior to the accretion region rotating into and out of view. The shape and one-magnitude amplitude of the pulse peak in this phase-averaged light curve both agree with the waveforms shown in \citet{nair05} and \citet{williams13}. Additionally, at the beginning and end of the orbital hump, there are brief, weak dips in the orbital waveform with a depth of $\sim$10\%; they have not been previously reported and are of unknown origin. A two-dimensional version of the orbital light curve (Fig.~\ref{fig:lightcurve}, lower panel) reveals that this waveform---including the dips---was extremely stable throughout C16, with no evidence of significant variations in the mass-transfer rate.


We applied the same analysis to the C18 data and found that the waveform was identical, with the exception of the dips. Due to their brevity and shallowness, they are likely undetectable at the thirty-minute cadence.

\section{discussion}


The accretion characteristics of EU~Cnc appear almost contradictory upon first glance. On one hand, there is an orbital variation of 1 magnitude in EU~Cnc, which is consistent with the findings of \citet{nair05} and \citet{williams13} and which suggests a relatively high mass-transfer rate. The latter study revealed that the orbital hump is due to cyclotron emission and that EU~Cnc is therefore actively accreting. However, the faint absolute magnitude of the system reported in \citet{nair05} is more consistent with a low-accretion state, as is its weak X-ray emission \citep[$\sim0.0005$ cnt s $^{-1}$ between 0.3-7 keV;][]{van04}. We interpret these observations as evidence that the orbital maxima of the light curve are attributable to magnetic accretion at a much lower mass-transfer rate than is normally observed in accreting polars. Additionally, there is an unusual consistency in the orbital light curve, with no dramatic change over the past 30 years of observation, suggesting that EU~Cnc has potentially remained in a low state across its observational history.

The low mass-transfer rate and consistency in the light curve suggest an alternate form of mass transfer than traditional Roche-lobe overflow, and many of the observational properties of EU Cnc could be explained if EU Cnc were a low-accretion-rate polar (LARP) in which accretion occurs via a magnetic siphon. As explained in \citet{schmidt05}, a magnetic siphon is a mechanism of mass transfer through which the secondary's entire stellar wind is captured, resulting in both weak X-ray emission and strong optical cyclotron harmonics, similar to those reported in \citet{williams13}. Nevertheless, the presence of weak He~I and He~II emission in spectra by \citet{pasquini} and \citet{williams13} suggests a higher accretion rate at times than what is typically observed in spectra of LARPs, which tend to show weak Balmer lines with a steep decrement and negligible He emission \citep{szkody}.

EU~Cnc therefore presents an intriguing challenge to the classification criteria for LARPs, and further observations of this system will help to clarify its nature. 


\end{document}